\documentclass[conference]{IEEEtran}
\IEEEoverridecommandlockouts
\usepackage{cite}
\usepackage{amsmath,amssymb,amsfonts}
\usepackage{algorithmic}
\usepackage{graphicx}
\usepackage{textcomp}
\usepackage{xcolor}
\usepackage{comment}
\usepackage{subcaption}
\usepackage{booktabs}

\def\BibTeX{{\rm B\kern-.05em{\sc i\kern-.025em b}\kern-.08em
    T\kern-.1667em\lower.7ex\hbox{E}\kern-.125emX}}

\definecolor{gray}{rgb}{0.4, 0.4, 0.4}

\usepackage{eso-pic}
\usepackage{url}
\AddToShipoutPictureBG*{
  \AtPageUpperLeft{%
    \put(0,-40){\raisebox{18pt}{\makebox[\paperwidth]{\begin{minipage}{21cm}\centering
      \textcolor{gray}{This article has been accepted for publicationn in the proceedings of the 9th ACM/IEEE Conference on Internet of Things Design and Implementation (IoTDI 2024).} 
    \end{minipage}}}}%
  }
  \AtPageLowerLeft{%
    \raisebox{20pt}{\makebox[\paperwidth]{\begin{minipage}{21cm}\centering
      \textcolor{gray}{ \copyright 2024 IEEE.  Personal use of this material is permitted.  Permission from IEEE must be obtained for all other uses, in any current or future media, including reprinting/republishing this material for advertising or promotional purposes, creating new collective works, for resale or redistribution to servers or lists, or reuse of any copyrighted component of this work in other works.
      }
    \end{minipage}}}%
  }
}

\begin{document}

\title{\emph{i}-CardiAx: Wearable IoT-Driven System for Early Sepsis Detection Through Long-Term Vital Sign Monitoring

}

\author{\IEEEauthorblockN{
Kanika Dheman\IEEEauthorrefmark{1},
Marco Giordano\IEEEauthorrefmark{1},
Cyriac Thomas\IEEEauthorrefmark{1},
Philipp Schilk\IEEEauthorrefmark{1},
Michele Magno\IEEEauthorrefmark{1}
}

\IEEEauthorblockA{\IEEEauthorrefmark{1}\{dhemank, mgiordano, cythomas, schilkp, magnom\}@ethz.ch}
\IEEEauthorblockA{\IEEEauthorrefmark{1}Center for Project Based Learning, Dept. of Information Technology and Electrical Engineering, ETH Z\"{u}rich, Switzerland}
}

\maketitle

\begin{abstract}
Sepsis is a major cause of premature mortality, high healthcare costs, and disability-adjusted life years. Digital interventions such as continuous cardiac monitoring solutions can help to monitor the patient's status and provide valuable feedback to clinicians to detect early warning signs and provide effective interventions. This paper presents \emph{i}-CardiAx, a wearable sensor based on low-power high-sensitivity accelerometers that measures vital signs essential for cardiovascular health monitoring, namely, heart rate (HR), blood pressure (BP), and respiratory rate (RR). A dataset has been collected from 10 healthy subjects with the \emph{i}-CardiAx wearable chest patch to develop low-complexity, lightweight vital sign measurement algorithms and evaluate their performance. The experimental evaluation demonstrates high-performance vital sign measurement for RR (-0.11$\pm$0.77 breaths per minute), HR ( 0.82$\pm$ 2.85 beats per minute), and systolic BP (-0.08 $\pm$ 6.245 mm of Hg). The proposed algorithms are embedded on the ARM Cortex-M33 processor supporting Bluetooth Low Energy (BLE). Estimation of HR and RR achieved an inference time of only 4.2 ms and 8.5 ms for BP. 
Moreover, a multi-channel quantized Temporal Convolutional Neural (TCN) Network has been proposed and trained on the open-source HiRID dataset to have a large number of patients with data for sepsis ground truth. The model has been trained and evaluated using only digitally acquired vital signs as input-data that could be collected by \emph{i}-CardiAx to detect the onset of Sepsis in a real-time scenario. The TCN has been fully quantized to 8-bit integers and deployed on \emph{i}-CardiAx.The network showed a median predicted time to sepsis of 8.2 hours with an energy per inference of 1.29mJ.
\emph{i}-CardiAx has a sleep power of 0.152 mW and an averages a power of 0.77 mW for always-on sensing and periodic on-board processing and BLE transmission. With a small 100 mAh battery, the operational longevity of the wearable has been estimated at two weeks (432 hours) for measuring the three cardiovascular parameters (HR, BP and RR) at a granularity of 30 measurements per hour per vital sign, running inference every 30 minutes. Thus, the wearable \emph{i}-CardiAx system can provide a method to monitor the cardiovascular parameters of patients with energy-efficient, high-sensitivity sensors to provide predictive alerts for life-threatening adverse events of sepsis, over a long period of time.
\end{abstract}

\begin{IEEEkeywords}
cardiovascular parameter monitoring, wearable, low power sensor nodes, continuous monitoring
\end{IEEEkeywords}

\section{Introduction}
Continuous monitoring of vital physiological parameters is crucial within intensive care units (ICUs), particularly for high-risk situations like premature birth or infants undergoing surgery. Sepsis, a potentially fatal syndrome resulting from infection-induced organ dysfunction, poses a significant burden in terms of morbidity and mortality, contributing to approximately 11 million annual deaths\cite{wentowski2021sepsis}. Its presentation varies across patients, making early identification challenging in different patient phenotypes\cite{fernando2018prognostic}. The prevalence of sepsis differs based on a country's economic demographic, but the global estimate stands at around 30 million cases each year, with a notable incidence among neonates and children, accounting for approximately 3 million and 1.2 million cases per year, respectively\cite{medicine2018crying}. Patient management is highly time-sensitive as delayed treatment increases the probability of mortality by 0.42$\%$ per hour of delay to administer antibiotics in sepsis patients \cite{makeranti}. Hence, identifying the onset of sepsis early and rapid initiation of antibiotics and supportive management are critical as no specific therapy is available to clinicians. \\

The current approach to sepsis screening primarily depends on monitoring vital signs and utilizing standardized scores such as the Systemic Inflammatory Response Syndrome (SIRS) score, Early Warning Score (EWS), and the Quick Sepsis-Related Organ Failure Assessment (qSOFA)\cite{vincent2016sepsis}. Nevertheless, these scoring systems are episodic and necessitate repeated measurements by medical personnel. Numerous factors impede effective sepsis monitoring of patients. Ongoing challenges in clinical implementation encompass the diverse clinical presentations of sepsis, the continuous and real-time tracking of vital signs, delays in accessing electronic medical records (EMR) due to third-party validation, a lack of medical decision support, alert fatigue, and information overload. In neonatal and pediatric ICUs, monitoring vital signs is further complicated by the presence of multiple hard-wired and inflexible connections to the delicate and developing skin of infants. These monitoring platforms pose significant risks of iatrogenic skin injuries, hinder skin-to-skin contact between the neonate and parent, complicate basic clinical tasks, and are incompatible with magnetic resonance and x-ray imaging.\\

In addition, pre-existing co-morbidities, such as cardiovascular disease (CVD) could also increase the risk of sepsis significantly \cite{arnautovic2018cardiovascular}. Unfortunately, non-communicable diseases (NCD) are the dominant health challenge in the twenty-first century and account for about two-thirds of all reported global deaths \cite{jagannathan2019global}. Cardiovascular diseases, in turn, account for 50\% of these global NCD-related deaths and are a major barrier to sustainable health development \cite{tsao2022heart}. The United Nations has set a goal under Sustainable Development Goal 3 to reduce premature mortality due to NCDs, and specifically CVDs by 30\% until 2030 \cite{lee2016transforming}. This necessitates the development and implementation of interventions that are cost-effective, unobtrusive, and comfortable to continuously monitor the major risk factors. In this regard, digital health solutions can provide an invaluable contribution to reducing adverse CVD outcomes, as shown by recent research \cite{widmer2015digital}. \\

To effectively track the onset of sepsis, particularly with a focus on monitoring CVD parameters, the most crucial vital signs are blood pressure, heart rate, and respiratory rate to prevent the occurrence of adverse events\cite{faletra2022assessing,brekke2019value}. Currently, wearable cardiac monitoring solutions, including Holter monitors, event recorders, ECG patches, wristbands, and smart textiles, predominantly offer heart rate measurements and electrocardiograms  \cite{widmer2015digital,vivalink}. These monitoring devices have limitations as they either do not provide all essential cardiovascular parameters (HR, RR, and BP) simultaneously, require multiple wires for frequently replaced electrodes, or have a design that hinders continuous usage\cite{duncker2021smart}. Additionally, certain high-risk populations, such as geriatric or paraplegic patients, require on-body alerts for sudden increases in cardiovascular parameters, especially systolic blood pressure \cite{krassioukov2009systematic}. However, the monitoring systems currently in use rely on external platforms for signal conditioning and communication to ensure accurate interpretation of bio-signals\cite{khan2016monitoring,yamamoto2017all}.\\

The ongoing IoT advancements, low-power sensors, and the integration of machine learning into low-power processors are transforming data collection, processing, and interpretation\cite{jiang2019lightweight,zainab2023lighteq}. This evolution is particularly significant for various applications, especially in the realm of wearables and medical devices, although it presents certain challenges \cite{SmartEMGWearable}. One major breakthrough is the use of low-power sensors, which optimize energy consumption, making them ideal for battery-operated wearables. By pairing these sensors with energy-efficient processors, the operational lifetime of devices is significantly extended. In-sensor data analytics are a key feature here, eliminating the need for additional processing hardware, ensuring clean data for clinicians and decision support systems, and enhancing device efficiency. Nonetheless, miniaturized and thin form factor wearables have strict constraints\cite{kallel2021critical}, including limited power\cite{schilk2023ear}, accuracy, and device longevity\cite{qaim2020towards}. Flexible hybrid sensors have been proposed as a solution but introduce potential reliability issues due to the integration of flexible and rigid electronics. Furthermore, the challenge of limited sensor lifespans due to battery dependence persists. Constant recharging and battery replacements are inconvenient and limit device capabilities. Therefore, addressing power challenges, whether through innovative power sources, energy harvesting, or more efficient processors, is crucial to fully unlock the potential of IoT in real-world applications, particularly in healthcare and beyond. In-sensor data analytics are crucial because they eliminate the need for additional hardware for processing and because they provide artifact-free information ready for interpretation by clinicians or medical decision support systems. Stringent constraints are set on the useable power, sensor accuracy, and lifetime of wearable devices due to the miniaturized and thin form factors. Flexible hybrid sensors with both flexible and rigid electronics have been proposed to circumvent such an issue, but these introduced reliability issues due to additional stresses in the electrical circuit\cite{luo2016flexible,yamamoto2016printed,khan2016flexible}.In this way, the lifetime of autonomous smart sensors is limited in all real-world applications by the use of batteries.\\

This paper proposes an energy-efficient and effective approach to assist in prognosis by continuously monitoring multiple cardiovascular vital signs to aid in disease management of sepsis patients by using only digital biomarkers. In particular, the paper presents the design and the implementation in hardware and software of emph{i}-CardiAx, a wearable system for continuously monitoring cardiovascular parameters of cuffless blood pressure, heart rate, and respiratory rate, medical decision support can be provided at the point of care using energy efficient multi-sensor smart patch with on-board tiny machine learning based algorithms in a mesh enabled IoT setup. The system is designed to be power efficient, hence, ensuring approximately 2 weeks of monitoring with small batteries with a maximum capacity of 100 mAh. To evaluate the capability of using only ultra-low-power sensors, the vital signs are extracted from two low-power accelerometers which are placed on the chest in a comfortable form factor for unobtrusive monitoring over long durations. The algorithms for vital signs measurement are designed to be low in complexity and run on a low-power ARM Cortex-M33 microcontroller (MCU).  Autonomous, low-power, and mesh-enabled multi-sensor wearable systems based on the always-on smart sensing paradigm can continuously acquire, process, and report physiological data in real-time and are trained to autonomously detect the onset of neonatal sepsis symptoms. Thus eliminating delays in electronic medical records and reducing alarm fatigue.\\

 The main contributions of this paper are:
\begin{itemize}
    \item  Design and development of an energy-efficient low-power sensing node based only on two accelerometers for cardiovascular parameter monitoring and onboard algorithm processing.
    \item Dataset creation of acceleration data on the chest at two different anatomical sites and experimental evaluation of vital signs: cuffless blood pressure (BP), heart rate (HR), and respiratory rate(RR), from bi-nodal IMU system.
    \item Development of deep learning model based on only cardiovascular vital sign extraction algorithms with low latency and memory requirements for implementation on the MCU.
    \item Experimental evaluation of the performance of proposed hardware and software solution, in terms of power consumption, energy, accuracy, and lifetime.

\end{itemize}
\begin{figure*}[hbt!]
    \centering
\includegraphics[width=0.9\textwidth]{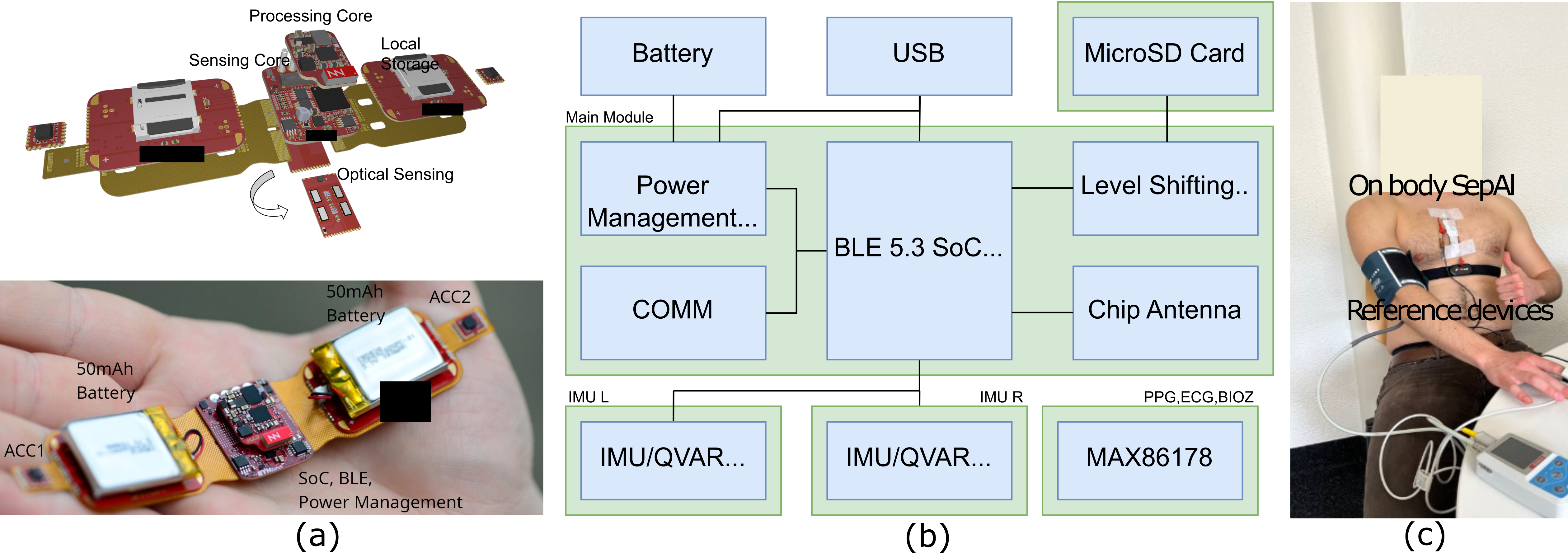}
   \caption{(a) \emph{i}-CardiAx wearable design showing the arrangement of the processing core, sensing core with the accelerometers on either extremity and the optical sensors at the bottom, (b) overview of the system hardware, and (c) \emph{i}-CardiAx system being tested on a subject with ground truth on vitals given by references devices }
    \label{fig:Sysarch}
\end{figure*}

\section{Related Work}
Research in predicting sepsis onset has leveraged data from electrical medical records using both statistical methods for patient survival analysis and deep learning models. Most statistical models assume a parametric function that subsumes the effect of input parameters to model a stochastic process. Such an assumption of a parametric relationship between the input and the hazard may not fully exemplify the heterogeneous nature of sepsis representation. Weilbull-Cox proportional hazards modeled 65 features extracted from the Emory University hospitals dataset and MIMIC-III dataset \cite{mimic}. Similarly, a Cox proportionality hazard model used 54 features to calculate a risk score for severe sepsis and septic shock\cite{henry2015targeted} with a prediction horizon of 28.2 hours before the onset of septic shock with a sensitivity of 0.85 and specificity of 0.67. Similarly many machine learning and deep learning algorithms on open-source electrical medical records from the ICUs have developed algorithms to predict sepsis onset. Temporal convolution networks (TCN) have been extensively applied to this task with one approach implementing a k-nn-based model with dynamic time warping that leveraged data sparsity by interpreting the missing data with a Gaussian process \cite{moor2019early}. 44 irregularly sampled laboratory and vital parameters were used for predicting the onset of sepsis 0 to 7h preceding sepsis onset. Another work used a TCN for sepsis prediction by addressing the missingness of data by masking the missing data in a parallel branch of the neural network instead of performing forward or back imputation \cite{multi_branch}. Gradient-boosted trees have also been used with the \textit{InSight} model where data from multi-center, multi-ward settings was labeled with the SIRS criteria at the University of California, San Francisco from 2011 to 2016 \cite{Calvert2016,Desautels2016,Shimabukuro2017}. This gave access to a large dataset with 90,353 patients for training on the model. This is orders of magnitude larger than any open-source EMR dataset and the result of this is seen in the high AUROC reported by the authors and reduced false positives \cite{Mao2018}. \\

However, most of these models use a large input vector dimension using almost all of the data in the electrical medical records. While the chemical bio-markers extracted at the laboratory are definitive in their prognosis, they are taken sporadically, are based on the analysis of the clinician, and require dedicated infrastructure and resources.  Also, the high input dimensions necessitate the use of dedicated computation infrastructure for providing medical decision support. All clinical environments may not be equipped with the required resources to provide such diagnosis, especially in low to middle-income countries (LMICs).\\

In contrast to many existing approaches that rely on extensive feature sets and resource-intensive models, our study takes a different direction. Our proposed approach has focused on creating a more efficient and accessible solution for predicting sepsis onset. Leveraging a carefully curated dataset and implementing a lightweight model, our approach seeks to address the limitations associated with high-dimensional input data derived from electronic medical records. While previous models have employed large input vectors that encompass a wide array of data from electronic medical records, we have prioritized a more streamlined approach. Our dataset selection process is meticulous, emphasizing data elements that are not only crucial for prognosis but are also available at a frequency that supports timely predictions. This approach aims to reduce the burden of dedicated infrastructure and resources required for sporadic chemical biomarker analysis, a common feature in many sepsis prediction models.\\

Recent advances in sensor technologies have led to many commercial monitoring wearables. However, most wearables focus on measuring the ECG and the HR from the chest. The Vivalink wearable ECG monitor provides multiple vitals such as HR and RR along with the ECG \cite{vivalink}. However, being an electrode-based system, it requires periodic electrode replacement and does not provide an estimate of the BP.

In such a scenario, low-power sensors such as accelerometers with high sensitivity are gaining popularity to provide solutions that allow long-term monitoring with low-complexity, lightweight algorithms that do not depend on the subject under test or the ambient conditions\cite{yin2021wearable,he2022smart,Chang2019}. The feasibility of measuring and co-relating the PTT with the BP was demonstrated with SCG on the carotid artery\cite{arathy2019accelerometric}. However, the placement of vibrational transducers on the neck can be excessively corrupted by speech, swallowing, and motion artifacts.

\begin{figure}[t]
    \centering
\includegraphics[width=0.5\textwidth]{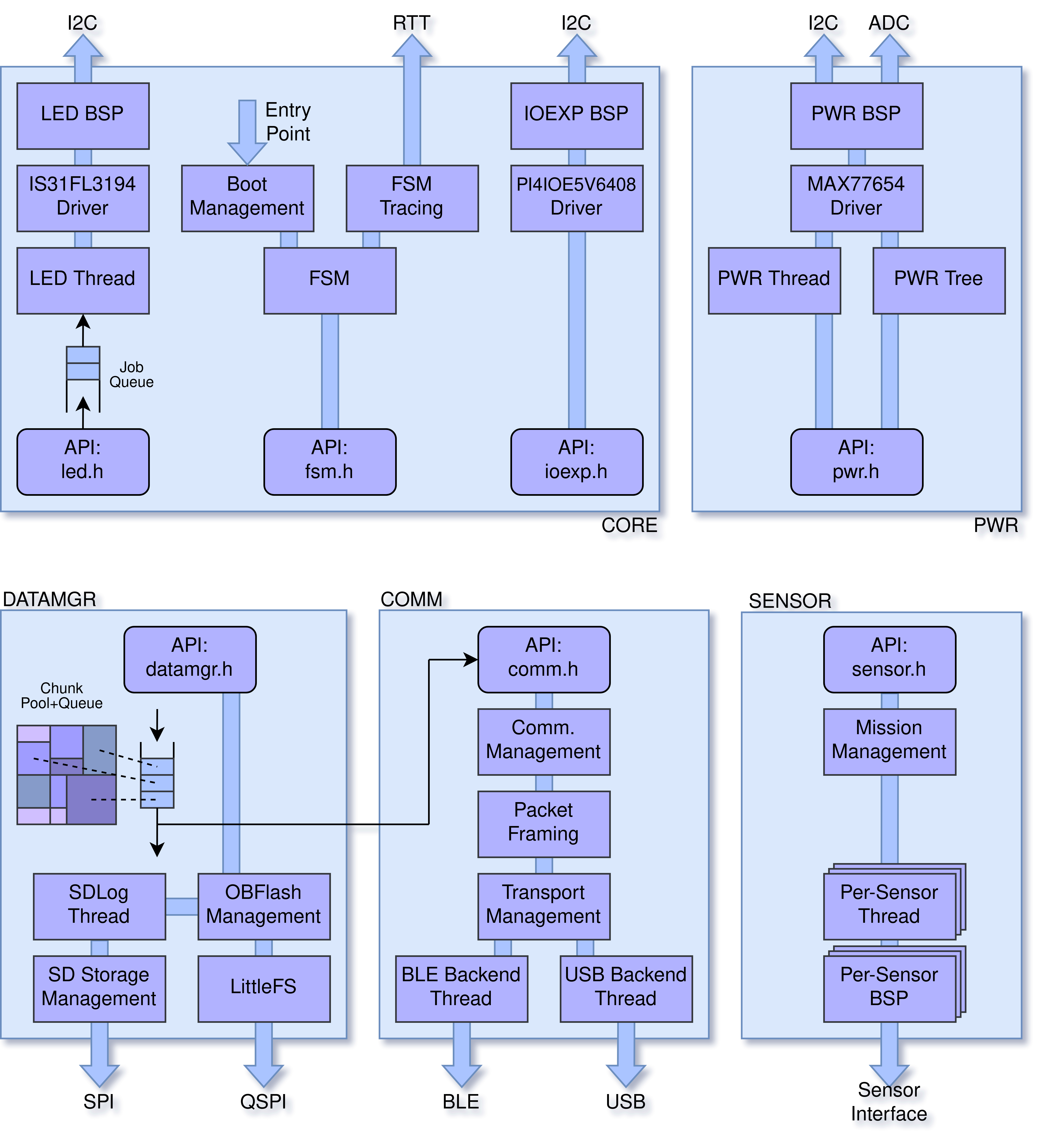}
   \caption{System operational overview showing the firmware architecture of the subsystems.}
    \label{fig:FW_arch}
\end{figure}

\begin{figure}[t]
	\centering
	\includegraphics[width=0.5\textwidth]{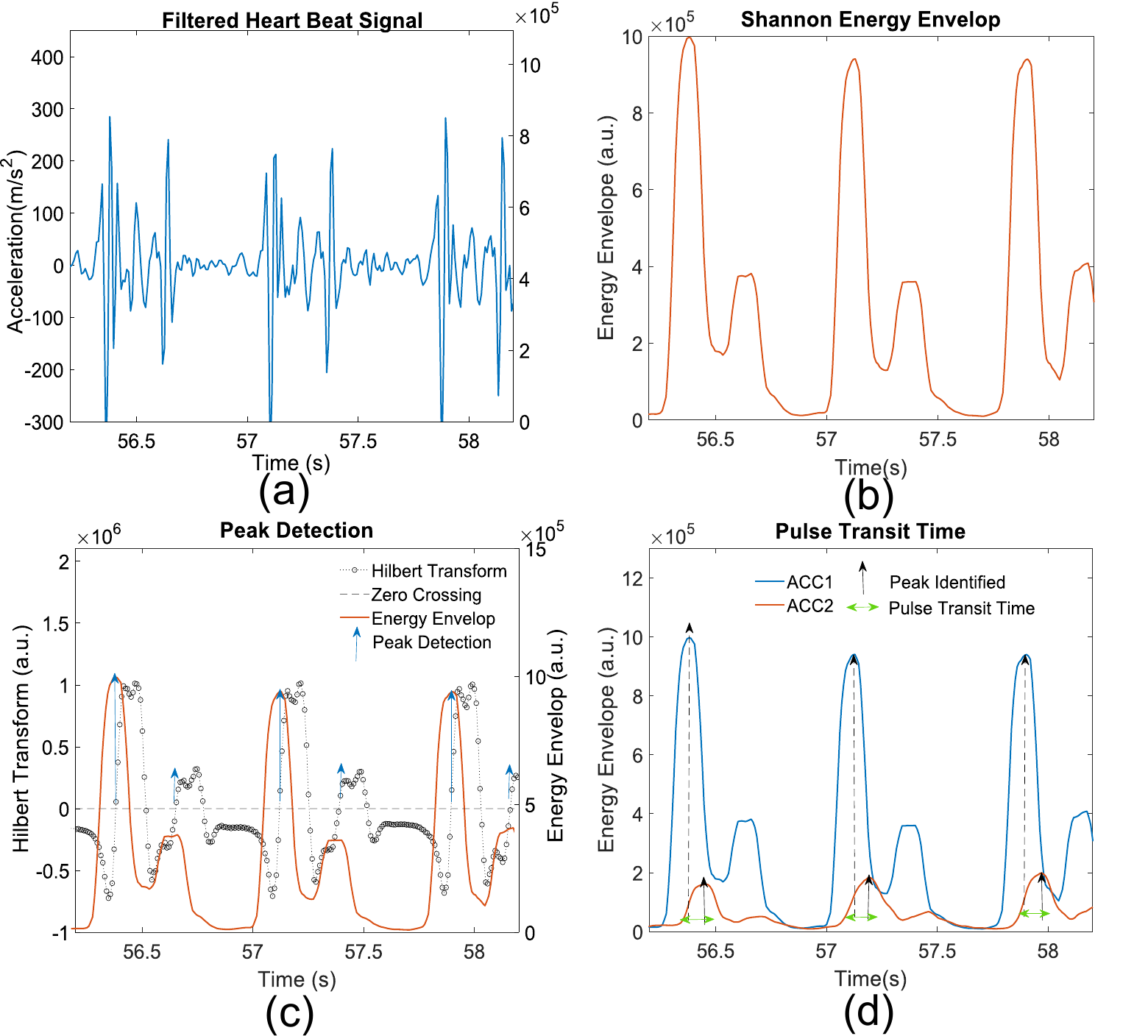}
    \vspace{-0.35cm}
	\caption{Algorithm for identification of heart peaks in the accelerometer signal with (a) bandpass filtering, (b) energy envelop of the vibrational signal, (c) peak identification with Hilbert transform and (d) calculation of the pulse transit time between the vibrational signal at the two anatomical sites.}
	\label{fig:algorithm}
    \vspace{-0.2cm}
\end{figure}

Cuffless BP estimation has been extensively pursued in research with photoplethysmography (PPG) and ECG binodal systems that attempt to capture the pulse transit time (PTT) from the ECG at the chest to the pulse arrival measured by the PPG at the fingertip \cite{kachuee2016cuffless}. Such systems are inconvenient for prolonged use and suffer the limitations of optical sensor-based PTT approaches, namely ambient light condition, skin morphology and proximity to skin \cite{mukkamala2015toward}. A promising approach is seismocardiography (SCG) using a 3-axis accelerometer which is independent of the individual under test and the ambient conditions to estimate the BP via vibrations on the chest\cite{Chang2019}. The feasibility of measuring and co-relating the PTT with the BP was demonstrated with SCG on the carotid artery\cite{arathy2019accelerometric}. However, the placement of vibrational transducers on the neck can be excessively corrupted by artifacts due to swallowing, talking, and head movements and the neck is a conspicuous position that intrudes upon the privacy of the user. \\

\begin{figure*}[hbt!]
    \centering
\includegraphics[width=0.8\textwidth]{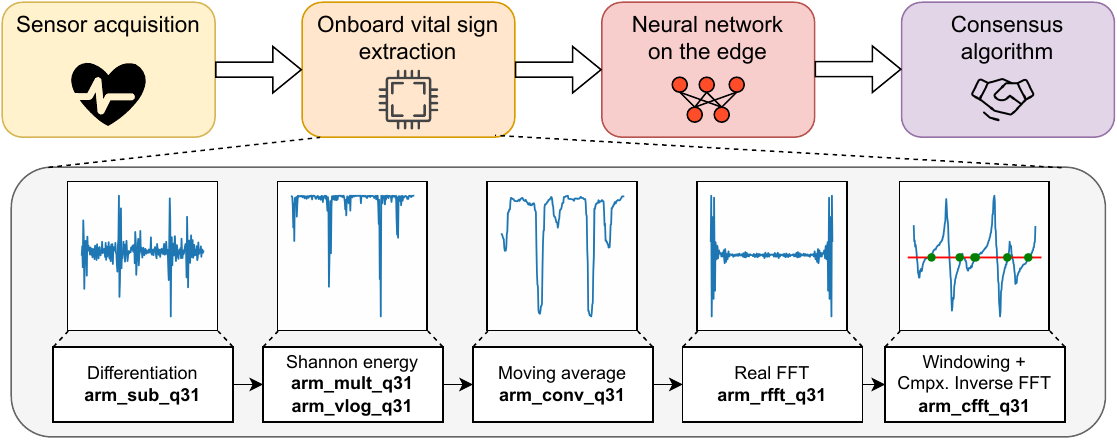}
   \caption{Data processing pipeline from sensor data acquisition to on-board vital sign extraction, which consists of multiple digital signal processing steps. After vital sign extraction, the vitals are fed to the neural network for prediction of sepsis onset.}
    \label{fig:vital_extraction}
\end{figure*}

This paper distinguishes itself from existing research in the field of blood pressure estimation in several significant ways. While the state-of-the-art has primarily focused on photoplethysmography (PPG) and ECG binodal systems to estimate blood pressure by capturing pulse transit time (PTT), our approach offers a fresh perspective using only two novel sensitive low-power accelerometers. In fact, many previous approaches rely on complex PPG and ECG systems that require sensor placement on both the chest and fingertip, making them inconvenient for prolonged use. Those sensors are susceptible to limitations associated with optical sensor-based PTT methods, including sensitivity to ambient light conditions, skin morphology, and sensor proximity to the skin. In contrast, this work presents an innovative approach using seismocardiography (SCG) with a 3-axis accelerometer. This method is not only independent of the individual under test and ambient conditions but also offers a more reliable means of estimating blood pressure through chest vibrations. Furthermore, while some prior research explored the feasibility of measuring and correlating PTT with blood pressure using SCG on the carotid artery, our work takes a different direction. We recognize the limitations of vibrational transducers placed on the neck, which can be prone to artifacts from swallowing, talking, and head movements. Additionally, this placement is conspicuous and may infringe upon the user's privacy. 

This paper endeavors to bridge this gap in wearable cardiac monitoring by presenting a multi-vital sign system that not only measures blood pressure but also heart rate (HR) and respiratory rate (RR). Beyond the accuracy of vital sign extraction, we delve into the implementation of on-device data processing, proposing a fully quantized temporal convolutional neural network, which enhances the system's usability and efficiency. Furthermore, we estimate the operational longevity of our compact wearable system, addressing a critical aspect of practicality and user-friendliness. In summary, our approach offers a more versatile and user-friendly solution to the challenges associated with blood pressure estimation and wearable cardiac monitoring.
 
\section{System Description}

\begin{figure*}
    \centering
    \includegraphics[width=0.8\textwidth]{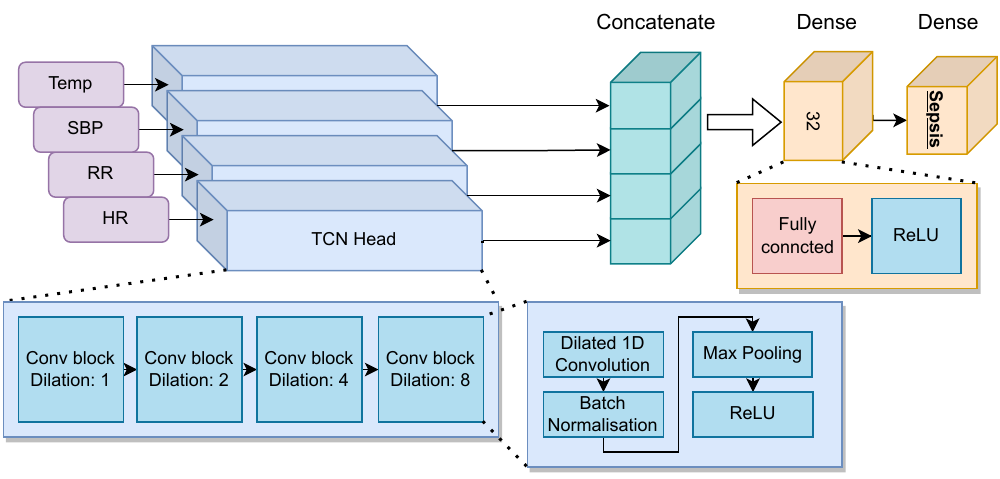}
   \caption{Neural network architecture showing the multi-TCN model with one TCN used per vital sign and a final concatenation in the final dense layer to identify a sepsis patient.}
   \label{fig:Neural Net}
\end{figure*}

\subsection{Hardware Overview}

At the heart of the \emph{i}-CardiAx is the BLE-enabled NRF5340 SoC which communicates and processes data. The SoC is clocked at 64 MHz and provides 512KB RAM, 1MB flash, and supports Bluetooth Low Energy (BLE). In addition, the system has the MAX77654 ultra-low power power-management IC that regulates the system power, an onboard flash, and 6-axis IMU (LSM6DSM). As seen in Figure \ref{fig:Sysarch}, these components form the main processing core of the \emph{i}-CardiAx and have been designed as a general-purpose controller for all subsystems. Using a high-density connector, this processing core can be attached to an additional circuit board featuring all sensors required for a given application and deployment. Attached on the sensing core, are both the optical emitters and sensors required for PPG sensing, a body temperature sensor, and two flexible interconnect-PCBs connecting to two side-wing PCBs, which in turn hold the batteries, micro-SD cards, and ECG-electrodes. The system is powered using 2 50mAh Li-ion batteries in parallel (total source capacity of 100 mAh). Finally, at the extreme ends of the device, are two high-performance always-on low-power LSM6DSV16BX inertial measurement units that can accurately measure the small vibrations related to blood pulse flow on the surface of the chest. The \emph{i}-CardiAx only uses the 3-axis accelerometers for extracting cardiac features on the chest at a sampling rate of 120Hz and a range of $\pm 2g$.

\subsection{Operational System Overview}

 The system-level overview is shown in Figure\ref{fig:FW_arch}. The main subsystems are:

\begin{itemize}
    \item A finite state machine (FSM) that controls the transition between different operating modes, such as idle, charging, and sensing. Transitions in the FSM are triggered by a predefined set of events that can be raised from all subsystems. Based on these events, the FSM will in turn control and configure all other subsystems using their public API It houses other components and peripherals that are used throughout the system, such as the 3-channel LED driver, IS31FL3194, and low-voltage translating 8-bit I2C-bus I/O ExpanderIO, PI4IOE5V6408.
    \item The PWR subsystem contains all power and resource management logic. It periodically monitors the battery charge level, charger insertion and removal, and battery temperature.
    \item The DATAMGR subsystem is responsible for collecting data produced by the onboard sensors and storing, transmitting, or retrieving it. All data storage in patchOS is built around protobufs\cite{Documentation}, which are compiled for this application using nanopb\cite{nanopb}. Every piece of data that needs to be logged or transmitted gets encapsulated and encoded as protobuf-defined data chunks allowing the DATAMGR subsystem to operate on variable-length blocks of binary data. This allows any changes to the data format to be made easily and without having to update any data management logic. For real-time transmission, these chunks are sent directly to the PC. For data logging, a collection of chunks is grouped into a so-called bundle, which is then stored in non-volatile storage.  Once the data is retrieved from local storage or streamed in real-time, the receiving side may easily decode the message using the protobuf definition and a protobuf library, making post-processing simple.
    \item The COMM subsystem enables Bluetooth and USB communication. This enables data to be offloaded, data to be streamed, and the operation of the device controlled.
    \item The SENSOR subsystem controls the management logic and interfaces to the different sensors available. Ultra-ow power, low noise (20 $\mu$g /$\sqrt{Hz}$) inertial measurement units, LSM6DSV16BX, provide a 6-axis accelerometer and gyroscope with charge variation (QVAR) capabilities along with a digital temperature sensor MAX30208.
\end{itemize}

\subsection{Low-Power Design}
\label{LP_Design}
The \emph{i}-CardiAx system was designed for low-power operation in order to ensure maximum operational longevity using low-power accelerometers (Active Power: 370 $\mu$W; Sleep Power: 12 $\mu$W ). The power-efficient system architecture is based on always-on low-power integrated circuits and energy-efficient power management. A real-time operating system (RTOS) handles the processing and control of the subsystems that allow the microcontroller unit (MCU) to stay in sleep mode as needed. To improve the energy efficiency the node has been designed to have a total sleep power of 0.155 mW. Experimental evaluation of the active power and sleep power will be illustrated in the following sections.  \\

Continuous operation mode is defined as the state where the sensors are always-on and the processor is woken only for two tasks: data collection from the FIFO and the processing of the vital sign algorithms. In continuous mode, the algorithms for HR and BP are executed every 2s, while the RR algorithm is executed every 30s. This frequency of tasks was heuristically chosen such that the accuracy of vital sign estimation is not compromised. At the end of every 30s, the average values of HR, BP, and RR are calculated. Hence, when the sensor is always on and vital signs are calculated in the order explained above, there are 30 measures (for the measured vitals of HR, RR, and BP) in an hour. Operational longevity can be further increased by reducing the number of measurements processed per hour and keeping the sensor node in sleep mode longer.

\subsection{Wearable Design}
The electronic layout was optimized for unobtrusive wearable applications. The main design objective was to record the heartbeat, respiration pattern, and transiting blood pulses. For this purpose, two accelerometers were used with one placed at the xiphoid process on the chest to capture the heartbeat and respiration patterns. The second accelerometer was placed at a distance of 12 cm, towards the suprasternal notch, from the first accelerometer to record the transiting blood pulse along the aortic pathway. This distance was chosen with the aim to capture the blood pulse from the xiphoid process to the aortic arch \cite{polar}. Figure \ref{fig:Sysarch} shows the wearable design with a compact form factor of 12 cm $\times$ 2.4 cm $\times$ 0.5 cm.
\begin{figure*}[t]
	\includegraphics[trim=0 2 5 90,clip,width=1.0\textwidth]{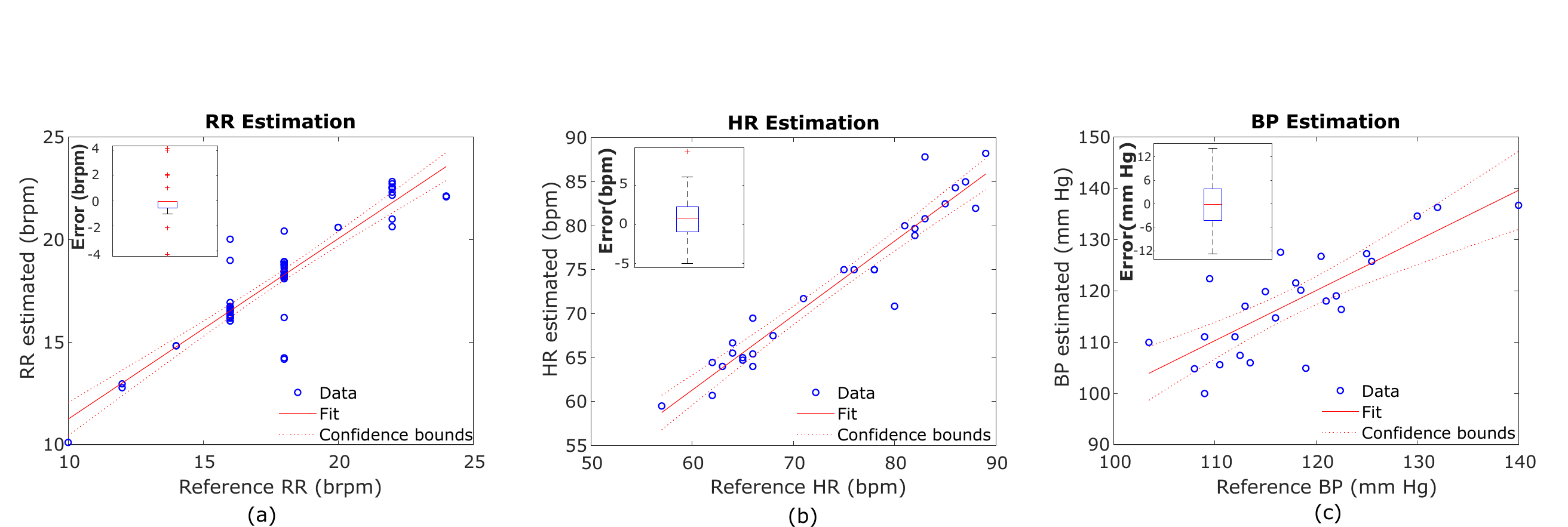}
	\caption{Estimation of cardiovascular vital signs: (a) respiration rate, (b) heart rate, and (c) systolic blood pressure as measured with the \emph{i}-CardiAx. Inset: error variation for each vital sign.}
	\label{fig:vital_performance}
    \vspace{-0.20cm}
\end{figure*}

\section{Methods}
Our primary objective is to demonstrate the feasibility of achieving sepsis alerts at the edge using only digital biomarkers, which can be extracted with two novel high-sensitive 3-axes accelerometers, while emphasizing the energy efficiency, and intelligence within \emph{i}-CardiAx. In the existing literature, studies that employed a restricted set of biomarkers for model training consistently incorporated the following vital signs: heart rate (HR), respiratory rate (RR), body temperature, and systolic blood pressure. \cite{Desautels2016,rapid_response}. By leveraging these digital biomarkers, \emph{i}-CardiAx offers a practical solution for sepsis detection without the need for invasive or additional sensors. Our system intelligently processes the data collected from these vital signs, enabling real-time analysis and timely alerts for sepsis onset. The careful selection and integration of low-power sensors ensure efficient power consumption, enabling prolonged device operation and minimizing the need for frequent battery replacements.

\subsection{Setup and Materials}
The Polar H10\cite{polar} was used as the reference device for HR and RR while the GIMA Ambulatory Blood Pressure Monitoring system was used as the reference for the blood pressure. Data was collected from 10 healthy subjects ( 5 males and 5 females) with the following demographics:  age of 25 $\pm 4.78$ years, weight in a range of 67 $\pm 7.97$  Kg, and height in the range of 172.8 $\pm 8.24$ cm.

\subsection{Experimental Protocol}
For the evaluation of vital signs, data was recorded in a sitting position. Subjects were asked to sit 5 minutes before starting the recording to stabilize the blood pressure due to hydro-static pressure changes from standing to sitting position.  The experiment was carried out in 3 segments of continuous measurements with the \emph{i}-CardiAx wearable for a period of 2 minutes each. All vitals of blood pressure, HR, and RR were measured before and after each segment. For the final comparison with ground truth, the mean of the estimates over a segment of 2 minutes was calculated to compare with the reference device measurement.

\subsection{Algorithms for Vital Sign Extraction}

\begin{itemize}
    \item \underline{Heart Rate (HR)}: The algorithm for extracting the peaks of the heart activity is shown in Figure \ref{fig:algorithm}. The Euclidean norm of the lateral and vertical axis is calculated which is then band-pass filtered between 10 Hz and 40 Hz (differentiation on the MCU). Thereafter, the Shannon energy envelope of the signal is calculated. The heartbeat peaks are located at the positive zero crossings of the Hilbert transform of the envelope, which is used to calculate the heart rate in windows of 2s. To evaluate the estimation error, the calculated average HR is compared to the reference device to report the average error and the variation of the error across subjects.   

    \item \underline{Cuffless Blood Pressure (BP)}: The same peak identification process, as described above for the HR estimation, is repeated for the two accelerometers. As seen in Figure \ref{fig:algorithm} (d), upon identification of peaks from both sensors, the pulse transit time is calculated as a difference of peaks located by ACC2 (12 cm away from the xiphoid process) and peaks located at ACC1 ( at the xiphoid process). The BP is evaluated for each subject from the mean PTT value over 2 minutes. This is due to the design of the experimental protocol, where the ground truth is taken from the cuff-based reference device before and after the 2-minute segment. For evaluation of the estimated blood pressure, a leave one out approach was used where \emph{n-1} subjects were used to extract the model of the blood pressure estimate using the PTT, and the error evaluation was done on the \emph{n$^{\mathrm{th}}$} subject. This process was repeated for all subjects to calculate the variation of the error for blood pressure estimation.

    \item \underline{Respiratory Rate (RR)}: Only the vertical z-axis is used for respiratory rate calculation over a window of 30s. The signal is downsampled to 15 Hz (8Hz on the MCU) to reduce the data size for computation. A Fast Fourier Transform is performed on this 30s data window to identify the maximum power in the frequency spectrum of 0.05Hz and 0.78 Hz to calculate the respiratory rate. On the MCU the same algorithm for HR extraction is used on the downsampled accelerometer signal for memory constraint reasons.

\end{itemize}

\begin{figure*}
    \centering
    \begin{minipage}{0.5\textwidth}
        \centering
        \includegraphics[width=0.8\textwidth]{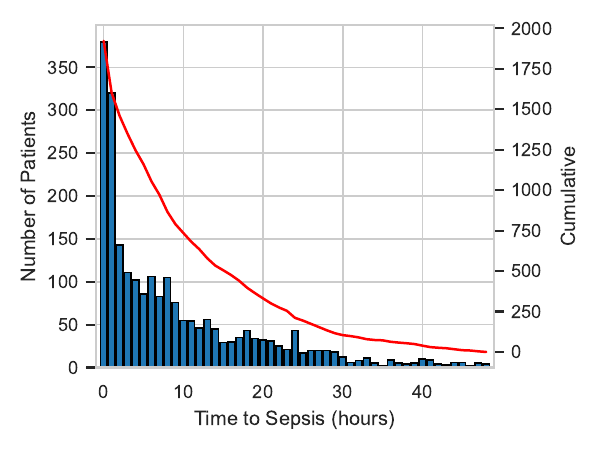}
        \subcaption{}
    \end{minipage}
    \begin{minipage}{0.49\textwidth}
        \centering
        \includegraphics[width=0.9\textwidth,trim=0 0.6cm 0 0,clip]{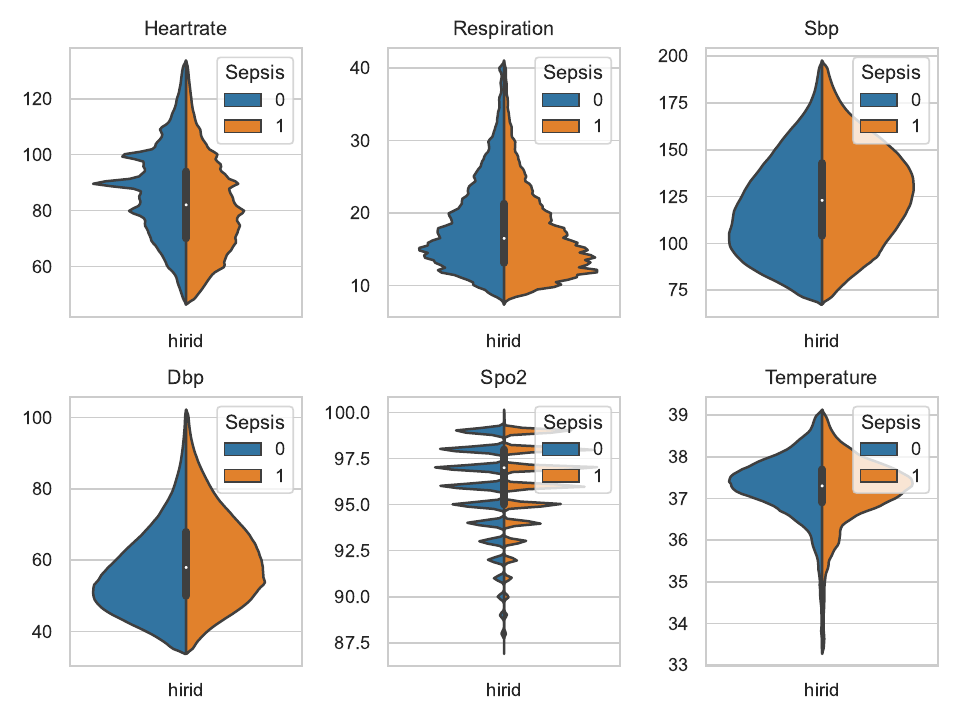}
        \subcaption{}
    \end{minipage}
   \caption{(a) Histogram of sepsis-positive patients in the HiRID dataset. (b) Violin plots of the distributions of the vital signs considered in this work: heart rate, respiration rate, systolic blood pressure (sbp), diastolic blood pressure (dbp), SpO2, and body temperature.}
   \label{fig:dataset_stats}
\end{figure*}

\begin{figure*}[htb]
    \centering
	\includegraphics[width=0.9\textwidth]{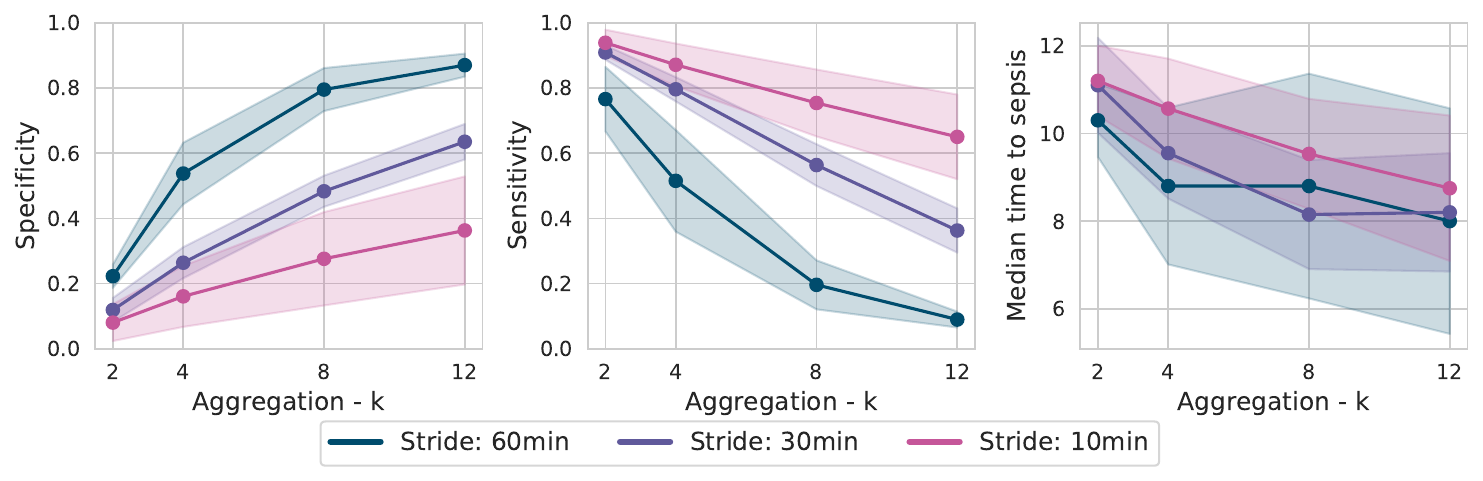}
	\caption{Specificity, sensitivity, and median time to sepsis onset for three window strides of 10 minutes, 30 minutes, and 60 minutes, against different aggregation parameters  \textsc{k}.  \textsc{k} is defined as the successive positive outputs of the TCN model which when aggregated together provide a consensus-based decision on a positive sepsis prediction.}
	\label{fig:Deeplearning_performance}
\end{figure*}

\subsection{On-board Processing}
The performance of the developed algorithm was evaluated on the ARM Cortex-M33 core on the \emph{i}-CardiAx system. The power consumption and execution time calculation of the algorithms on the proposed device was tested with a Power Profiler Kit II from Nordic Semiconductors\cite{Nordic_Semiconductor}. \\

Figure \ref{fig:vital_extraction} shows the signal processing pipeline and the optimized CMSIS-DSP functions implemented on the ARM Cortex-M33 microcontroller on board of \emph{i}-CardiAx. The differentiation is obtained with a subtraction operation, the Shannon energy with the multiplication of the power, and the logarithm of the power of the differentiated signal. Then a smoothing moving average is implemented with a convolutional filter before extracting the Hilbert Transform. The Hilbert Transform is obtained by convolving a windowing function to the smoothened Shannon energy, which is implemented by extracting the frequency components of the signal, windowing them, and taking the imaginary component of the inverse transform of the obtained signal. This last step is shown in Figure \ref{fig:vital_extraction}, where the zero crossings, corresponding to the peak positions, are highlighted in green.\\

Blood pressure is extracted by taking the pulse transit time between the identified peaks of the two accelerometers which capture the heartbeat at the two extremities. A model between the pulse transit time and the ground truth BP is deduced to estimate the BP values.

\subsection{ On-device Model Architecture for Sepsis Onset Detection}
\label{section: Model_TCN_arch}
Building on top of \cite{10599159}, we integrated a Temporal Convolutional Network (TCN) architecture for sepsis onset detection in our prototype.
Currently, sepsis detection relies on the comprehensive analysis of various parameters in electronic medical records. Nevertheless, the prospect of ubiquitous sepsis monitoring via wearable devices at the point of care hinges on the ability to identify sepsis onset using digital health biomarkers, particularly vital signs. To determine the feasibility of recognizing sepsis onset exclusively through vital signs, we constructed a retrospective model using solely digitally captured vital signs from electronic medical records. Subsequently, we developed and assessed a real-time sepsis onset detection model, aiming to identify sepsis occurrences at specific intervals.

\subsubsection{Dataset}
To train and test a neural network to identify sepsis onset early, an open-source medical data set, HiRID was used \cite{yeche2021hirid}. The HiRID dataset comprises 34 thousand patient records from the Department of Intensive Care Medicine at Bern University Hospital in Switzerland (ICU). It includes anonymized demographic information, real-time measurements from bedside monitors, usage of medical devices like mechanical ventilation, observation notes by healthcare providers, laboratory-acquired biochemical markers, administered drugs, fluids, and nutrition .\\

\begin{figure*}[h]
    \centering
	\includegraphics[width=0.9\textwidth]{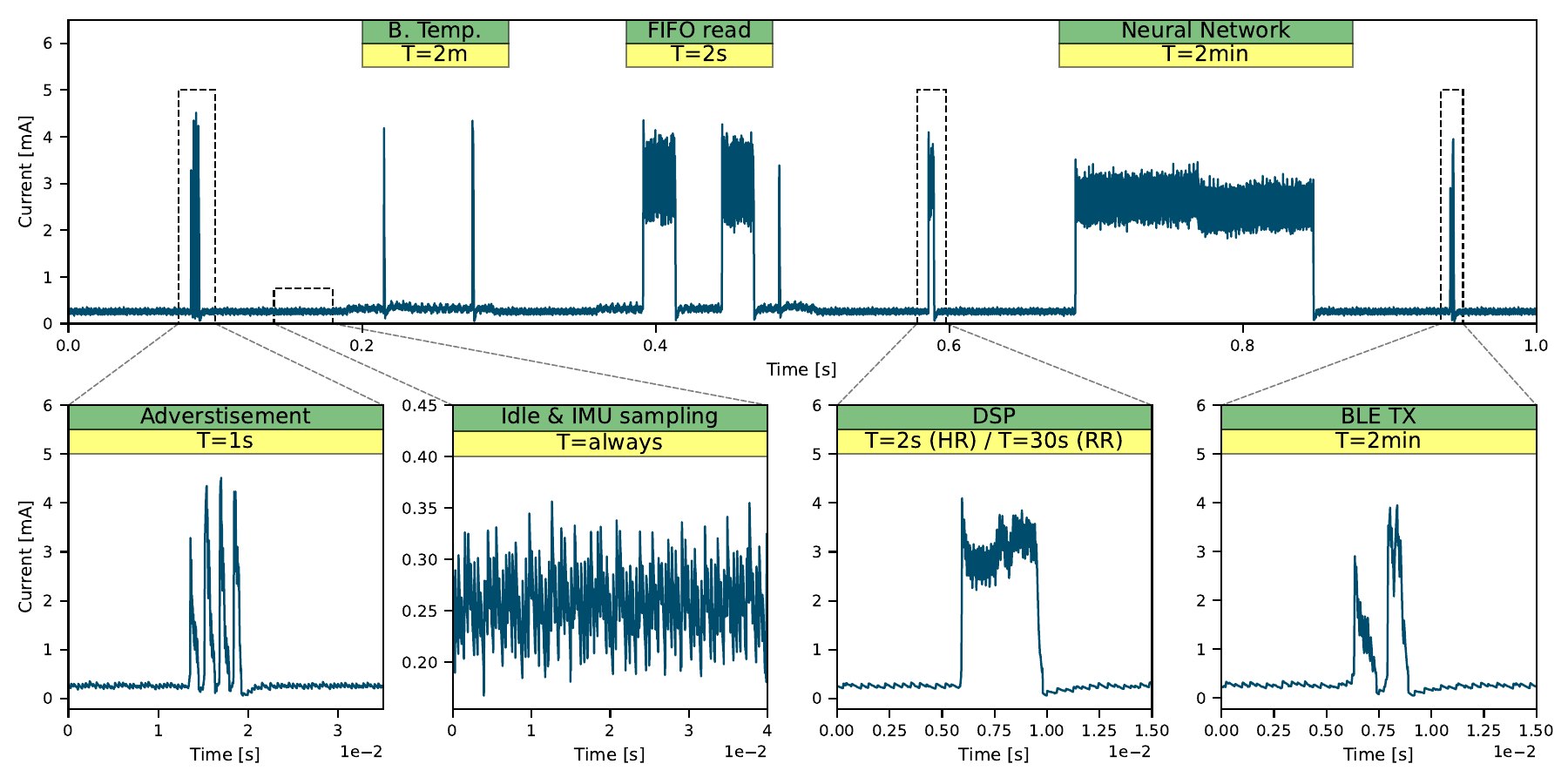}
	\caption{Power profiling of the system. The different tasks are reported with execution period T.}
	\label{fig:power_profiling}
    \vspace{-0.35cm}
\end{figure*}

\subsubsection{Inclusion Criteria and sepsis-onset labeling} Patient selection adhered to specific inclusion criteria in line with prior research. These criteria encompassed a minimum hospital stay of 24 hours, a minimum age of 18 years, exclusion of patients who received antibiotics within the initial 7 hours of admission to the ICU, and exclusion of patients with a time to sepsis onset of less than 4 hours (as sufficient time is needed to identify patterns and thresholds by the machine learning algorithms). To ensure a balanced representation of positive and negative cases during training, the maximum duration of ICU stay was limited to 48 hours. Sepsis onset timing was determined in accordance with the sepsis-3 criteria \cite{singer2016third}. A window for suspicion of infection was identified by pinpointing the time of antibiotic administration and creating a time window extending 48 hours before and 24 hours after this moment. For a patient, sepsis onset was determined when the hourly change in the Sequential Organ Failure Assessment (SOFA) score was greater than or equal to 2. It was assumed that sepsis-specific features would be present from the outset, so each 4-hour data window was labeled as class 1 for sepsis-positive cases and class 0 for all other patients. Following the application of these criteria, the dataset consisted of 1058 sepsis-positive patients and 7635 sepsis-negative patients, resulting in a class imbalance of 13.9\%. Furthermore, to enable real-time prediction of risk leads to labelling each time window of 1 hour with a label of 1 for a positive class and a label of 0 for a control case. The data was split into training and test sets with an 80/20 ratio. Importantly, only the training set was balanced, ensuring the test set remained unaltered for a fair comparison under real-life conditions.

Figure \ref{fig:dataset_stats}(a) illustrates the distribution of sepsis-positive patients concerning the time to sepsis onset. The histogram reveals that over half of sepsis-positive patients develop sepsis within the first 8 hours of their ICU stay. This presents a challenge due to the limited data available for training machine learning models and exacerbates the class imbalance. Figure \ref{fig:dataset_stats}(b)  shows the distribution plots of the vital signs in the EMR that can be digitally acquired via wearable devices for both sepsis-positive and negative patients. Notably, some vital signs, such as blood pressure and heart rate, exhibit significant differences in distribution between the two groups.

\subsubsection{Temporal convolution network (TCN)}
 A multi-head Temporal Convolutional Network (TCN) has been proposed and designed with a focus on implementation on low-power microcontrollers. TCNs can identify causality in time series data and are well-suited for real-time sepsis detection \cite{bai2018empirical,9765481}. The TCN architecture, illustrated in Figure \ref{fig:vital_extraction}, differs from previous sepsis detection models utilizing TCN architectures \cite{moor2019early,multi_branch,Rosnati2021} in that it leverages the causal relationships within each vital sign individually with a singular TCN at first which is followed by concatenating and flattening the outputs from all 'n' TCNs in the final fully connected layer. such a  design aims to learn multivariate relationships in the final dense layers, thus enabling the model to learn distinctive features from each input stream.\\

 As shown in Figure \ref{fig:Neural Net}, the TCN uses only digitally acquired vital signs. Model optimization with respect to model hyper-parameters, such as the number of layers, convolution filters, learning rate, and schedule, a weight-informed Neural Architecture Search (NAS) was utilized \cite{ren2021comprehensive}. The optimized architecture features 4 layers of dilated causal convolution, with a power-of-2 exponential increase in dilation size to capture longer-range dependencies in time-series data, along with 32 filters in each layer. The output of the convolution layer is flattened and passed into a dense layer comprising 32 neurons. Each convolution layer has a kernel dimension of 3, and batch normalization follows each convolution layer. A max pooling layer with a kernel and stride of 2 and a ReLU activation function is applied. The final layer employs a Sigmoid function, providing an interpretable output as a probability.

\subsubsection{Input data vector}
The input of the model consists of 4 hours of data sampled at 2 samples/minute using the high-granularity HiRID dataset. Digital biomarkers from the EMR that were considered while developing the model are the heart rate, systolic blood pressure, respiratory rate, and core body temperature. The selection is motivated by the feasibility of acquiring these markers with onboard sensors, such as low-power commercial IMUs \cite{magno2019self}, making the system non-dependant on external laboratories and/or manual data entries.

\subsubsection{Sepsis prediction with a consensus algorithm} A consensus algorithm was applied to the output of the TCN over successive windows for real-time identification of sepsis onset, as depicted in Figure \ref{fig:vital_extraction}(d). When \textsc{k} successive windows are labeled as 1, the alarm is raised and the patient is labeled as sepsis positive. An analysis of possible values of \textsc{k}  and window stride is given in Section \label{sec:exp}.

\begin{table}[t]
\centering
\begin{tabular}{lcc}
\toprule
Metric      & Float & INT8 \\
\midrule
Sensitivity  & 0.56          & 0.52   \\
Specificity  & 0.48           & 0.54  \\
Med. t. sep. & 8.2 h          & 8.1 h   \\
\bottomrule
\end{tabular}
\caption{Performance comparison between floating point and quantized model.}
\label{tab:quant_float_comp}
\vspace{-0.5cm}
\end{table}

\begin{figure}[t]
	\centering
	\includegraphics[width=0.5\textwidth]{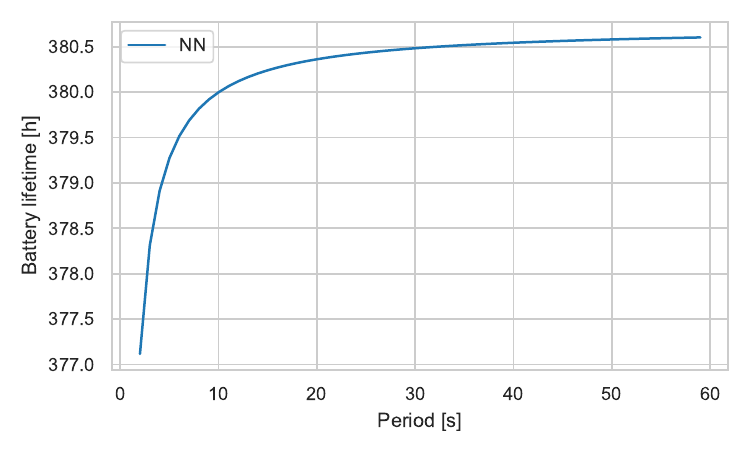}
	\caption{Battery lifetime estimate varying the neural network window stride.}
	\label{fig:battery_estimation}
 \vspace{-0.8cm}
\end{figure}

\section{Experimental Results}
\label{sec:exp}

\subsection{Performance: Vital Sign Estimation Accuracy}
The system performance in terms of correlation with the reference device and the variation of estimation error for the measured vital signs is shown in Figure \ref{fig:vital_performance}. RR has an R$^{\mathrm{2}}$= 0.82 with an error variation for RR estimation of -0.119 $\pm$ 0.7 breaths per minute (brpm). The error variation for HR is 0.829 $\pm$ 2.8 beats per minute (bpm) and a high correlation with R$^{\mathrm{2}}$= 0.92. The BP measure for the systolic blood pressure shows a correlation with R$^{\mathrm{2}}$= 0.67. the average error in BP estimation is -0.0849 mm of Hg and the variance of the error is $\pm$ 6.24 mm of Hg, thereby fulfilling the Association for the Advancement of Medical Instrumentation (AAMI) criteria where the tolerable error is less than 10mm of Hg \cite{topouchian2022accuracy}.

\begin{table*}[]
\centering
\begin{minipage}{0.37\textwidth}
\centering
\begin{tabular}{cl}
\toprule
\textbf{Sub-system} & \textbf{Power} \\
\midrule
nRF5340 Sleep & 88 $\mu$W  \\
MAX77654 (Quiescent draw)& 58 $\mu$W  \\
LSM6DSV16BX Active$^{\mathrm{a}}$ & 370 $\mu$W  \\
LSM6DSV16BX Sleep$^{\mathrm{a}}$ & 6 $\mu$W  \\
\multicolumn{2}{l}{$^{\mathrm{a}}$ 3-Axis accelerometer only.}
\end{tabular}
\caption{Per component power profiling}
\label{tab:low_power}
\end{minipage}
\begin{minipage}{0.37\textwidth}
\centering
\begin{tabular}{clll}
\toprule
Task  & Period & Power {[}mW{]} & Energy {[}uJ{]} \\
\midrule
Idle  & -      & 0.77           & -               \\
FIFO  & 2s     & 5.32           & 144.6             \\
B. Temp  & 2m     & 1.08           & 108             \\
DSP HR& 2s     & 5.22           & 18.8            \\
DSP RR& 30s     & 5.22           & 18.8            \\
NN    & 2m     & 6.46           & 1.29$\cdot 10^3$       \\
Adv.  & 1s     & 2.8           & 14            \\
TX 1B & 2m    & 1.39           & 20.9           \\
\bottomrule
\end{tabular}
\caption{Per task power profiling}
\label{tab:tasks}
\end{minipage}
\begin{minipage}{0.25\textwidth}
\centering
\begin{tabular}{cl}
\toprule
\textbf{Sub-system} & \textbf{Power} \\
\midrule
NN & 377h \\
Streaming & 180h \\
\end{tabular}
\caption{Battery lifetime estimation}
\label{tab:battery}
\end{minipage}
\end{table*}

\subsection{On-Board Deep Learning Model Performance}
Figure \ref{fig:Deeplearning_performance} presents the model's specificity, sensitivity, and median time to sepsis predictions for various values of the aggregation parameter (\textsc{k}) and three different stride intervals of 10 minutes, 30 minutes, and 60 minutes. The solid line represents the mean, while the shaded area reflects the standard deviation over a 5-fold cross-validation performed on the test dataset.\\

An increased overlap between consecutive windows when combined with a shorter stride introduces redundancy which yields a low true positive rate as compared to longer window strides. On the other hand, longer windows may miss the causal dependencies between sepsis-informative features, resulting in reduced specificity and an elevated false positive rate. Increasing \textsc{k} makes the consensus algorithm more reliant on consecutive positively predicted windows, enhancing specificity. However, this increased reliance also necessitates more positive predictions to trigger an alarm, consequently elevating the false negative rate. The median time to sepsis was calculated by using stride of 30 minutes and an aggregation parameter of 8. This parameter selection resulted in a sensitivity of (0.56 $\pm$ 0.06) and specificity of (0.48 $\pm$ 0.047) and an estimate of median predicted time to sepsis of 8.2 hours (CI: [7 h, 9.4 h]).

The performance of the TCN is hampered by the confinement to the HiRID dataset, which lacks the diversity of a larger sample. However, this dataset was chosen due to its high granularity of vital sign measurements and being the only open-source dataset available.\\

In Table \ref{tab:quant_float_comp} the results for the quantized model are reported. It can be seen that the performance difference between the floating point and the quantized model is below 10\%, with a 4x memory usage improvement.

\subsection{On-board Implementation of vital sign algorithms}
The onboard implementation on the ARM Cortex-M33 was clocked at 64 MHz.The HR and RR peak detection took 4.05 ms with a memory footprint of only 38 kB, including 2s of accelerometer data and algorithm parameters. The BP algorithm used the already identified peaks in the first task from one sensor (ACC1) and then executed the peak detection again on the second accelerometer (ACC2) to calculate the PTT, yielding the same metrics reported before. 1.4 ms were taken to calculate the average values of the HR and BP. The RR algorithm is executed every 30s and has the same metrics as the HR algorithm, but a lower filtering frequency to accommodate the slower-evolving respiratory signal.

\subsection{Power Profiling}
 In Figure \ref{fig:power_profiling} the power profiling for a standard workflow for \emph{i}-CardiAx is reported. First, the node enables BLE advertisement packets, while simultaneously collecting data from the sensors. Once the FIFO buffer inside the sensors is full, every 2 seconds, the data is read by the MCU which is followed by vital sign extraction (digital signal processing (DSP) block as shown in Figure \ref{fig:vital_extraction}). The TCN inference is run after the TCN-input buffer of 4 hours is filled. Then for every time window stride, an inference is executed. The prediction of the TCN is saved for every inference and fed into the consensus algorithm, which when it encounters the conditions that satisfy the triggering of an alarm, transmits the alarm via BLE. 

In Table \ref{tab:low_power} the current consumption and the latency of each state are reported, while in Table \ref{tab:tasks} an overview of the active tasks is shown. The IDLE state of the system is defined as all subsystems in sleep mode with BLE advertising every 1s and the IMU collecting accelerometer data at 120Hz. Data is fetched from the sensors every 2 seconds with an average power consumption of 5.32 mW and takes 27 ms for the MCU to execute. The vital sign extraction is performed every 2s to extract the peaks from the accelerometer data. The total current consumption for this is 5.22 mW throughout 3.6 ms per accelerometer. 
The TCN inference is energy efficient and consumes 6.49 mW over 320 ms and the task is run every 30 minutes on a full NN-input buffer and according to the window stride. The BLE transmission of the inference accounts for 1 byte that is transferred in 0.5 ms at a power consumption of 1.39 mW. The operation of \emph{i}-CardiAx is lightweight (approximately 100 kB of memory footprint) and always within mW-envelopes

\subsection{Battery Lifetime}
Based on the power requirements of the sensors, onboard processing of vitals, BLE transmission, power supply quiescent draws, and other system management SoC tasks, a battery lifetime for different rates of on-board vital sign inference is estimated in Figure \ref{fig:battery_estimation} and summarised in Table \ref{tab:battery}. The estimated operational longevity is 337 hours with a neural network window stride of 2min and up to 455 with a neural network window stride of 60 minutes. For the case analyzed in this work (window stride of 30min), the estimated battery lifetime is 432h, around 18 days, with a battery of 100mAh. The simulation is done under the task period reported in Table \ref{tab:tasks}
\section{Conclusions}
 This paper presented \emph{i}-CardiAx, a long-lasting wearable system for continuous cardiovascular parameter monitoring using only two high-sensitive low-power accelerometers. The developed algorithms for HR, BP, and RR showed high accuracy in vital sign estimation when compared with reference devices and were compliant with international standards for medical instrumentation, such as AAMI for blood pressure monitoring. Additionally, the algorithms were fully embedded in a low-power ARM Cortex-M33  and transmitted via BLE. Due to the onboard processing and the low-power design, we demonstrated that  \emph{i}-CardiAx can last for approximately two weeks with a single 100mAh battery when the sensor is always on and vitals of HR, RR, and BP are reported every 30s.


\bibliographystyle{ieeetr}
\bibliography{bsn2023}

\end{document}